\documentclass[reprint,amssymb, amsmath, aip,cha,twocolumn]{revtex4-1}
\usepackage{graphicx}
\usepackage[caption = false]{subfig}
\usepackage{color}
\usepackage{epsfig}
\usepackage{bm}%
\usepackage[colorlinks=true,linkcolor=blue]{hyperref}%
\expandafter\ifx\csname package@font\endcsname\relax\else
 \expandafter\expandafter
 \expandafter\usepackage
 \expandafter\expandafter
 \expandafter{\csname package@font\endcsname}%
\fi
\begin{document}
\title{Experimental investigation of flow induced dust acoustic shock waves in a complex plasma}%
\author{S. Jaiswal}%
\email{surabhijaiswal73@gmail.com}
\author{P. Bandyopadhyay}
\author{A. Sen}
\affiliation{Institute For Plasma Research, Bhat, Gandhinagar,Gujarat, India, 382428}%
\date{\today}
\begin{abstract}
We report on experimental observations of flow induced large amplitude dust-acoustic shock waves (DASW) in a complex plasma. The experiments have been carried out in a $\Pi$ shaped DC glow discharge experimental device using kaolin particles as the dust component in a background of Argon plasma. A strong supersonic flow of the dust fluid is induced by adjusting the pumping speed and neutral gas flow into the device. An isolated copper wire mounted on the cathode acts as a potential barrier to the flow of dust particles. A sudden change of gas flow rate is used to trigger the onset of high velocity dust acoustic shocks whose dynamics are captured by fast video pictures of the evolving structures. The physical characteristics of these shocks are delineated through a parametric scan of their dynamical properties over a range of flow speeds and potential hill heights. The observed evolution of the shock waves and their propagation characteristics are found to compare well with model numerical results based on a modified Korteweg-de-Vries-Burgers type equation. 
\end{abstract}
\maketitle
\section{Introduction}\label{sec:intro}
A dusty (or complex) plasma consists of a suspension of small (nanometer to micrometer sized) dust grains in a conventional two component electron-ion plasma. These heavy dust particles embedded in the plasma environment normally become negatively charged by collecting more electrons than ions. In laboratory devices that employ electrodes to generate the plasma the dust particles levitate in the boundary region of the plasma sheath formed at the cathode. An equilibrium dust layer results from a balance between the forces exerted by the sheath electric field (pulling the dust upwards) and gravity (pulling the dust downwards). Such an equilibium when perturbed can give rise to collective oscillations in the dust cloud. Due to the small charge to mass ratio of the dust particles compared to that of electrons and ions, the time scales of dust dynamics are comparatively much longer and easier to track visually. Thus dusty plasmas have provided an excellent medium for investigating such processes as phase transitions and also for exploring collective excitations in its gaseous, liquid and solid state manifestations \cite{Ikezi, thomas_1994}. There is a rich literature available on the study of linear and non-linear modes and coherent structures in a complex plasma such as dust-ion-acoustic (DIA) waves
\cite{ma}, dust-acoustic (DA) waves \cite{rao}, dust lattice (DL) waves \cite{melands}, dust Coulomb waves \cite{N}, dust voids \cite{goree} and vortices \cite{law}. Among them, the DIA solitary waves, the DA solitary waves, the DL solitary waves, Dust ion-acoustic shock waves (DIASW)\cite{bailung} and Dust-acoustic shock waves (DASW) \cite{Andersen} are very important nonlinear waves which have been extensively studied both theoretically \cite{shuk, kour, morf, versheest, mamun}, and experimentally \cite{Pintu, samsonov, heidemann} over the past couple of decades.\par
Shock waves constitute a special class of nonlinear waves in the form of propagating discontinuous disturbances that are characterized by sharp jumps in velocity, pressure, temperature and density across a narrow front. They occur in a variety of neutral media such as in gases \cite{j.bond.1965}, fluids \cite{dimon 1999} and solids \cite{R. Graham1993}. They have also observed in space during the early stages of strar fomation and the interaction of the magnetic fields of Earth and the solar wind. In  plasma medium, they can be excited  when a large amplitude mode propagates in the  presence of strong dissipation such as due to collisions with neutrals, viscosity or Landau damping\cite{sagdeev, andersen 1968}. The dynamics of shock waves have also been studied in complex (dusty) plasmas both theoretically as well as experimentally. In a weakly coupled dusty plasma kinematic viscosity\cite{shukla2000} arising from dust-ion collisions and the presence of dust charge fluctuations \cite{popel2000} have been identified as two important dissipation sources contributing to the formation of shock waves. In a strongly coupled dusty plasma, on the other hand, the shear and bulk viscosities play an important role in the formation of  monotonic and oscillatory dust acoustic shock waves \cite{shukla2001, hamid2009}. Samsonov\textit{et al.}\cite{samsonov2003} were the first to report an experimental observation of shock formation in a rf produced 3D complex plasma under  microgravity conditions in the PKE-Nefedov device. In that experiment, a sudden gas pulse from an electromagnetic valve was used to excite the shock waves. Soon after, Fortov \textit{et al.}\cite{fortov2005} reported the propagation of compressional shock waves in a dusty plasma in which an impulse of axial magnetic field was used to excite dust acoustic shock fronts. In 2009, Heinrich\textit{et al.}\cite{heinrich2009} observed the repeated occurence of self-excited dust acoustic shock waves (DASWs) in a DC glow discharge dusty plasma. The shocks were generated when the dust cloud went through two slits separated by a very small distance.  Nakamura \textit{et al.}\cite{nakamura2012} in 2012, reported the experimental observation of a bow shock like formation in a two-dimensional complex plasma due to the super-sonic flow of charged microparticles through a stationary object. More recently, Usachev \textit{et al.}\cite{usachev2014} have experimentally investigated the formation and dissipation of an externally excited planar dust acoustic shock wave in a three-dimensional uniform dust cloud under the microgravity conditions.\par
 
To the best of our knowledge, there has not been any detailed experimental observation of flow induced excitation of dust acoustic shock waves and its propagation characteristic with the different perturbation strength in terms of flow velocity and hill height. In this paper, we have experimentally investigated neutral flow driven excitation and propagation characteristics of shock waves in a DC glow discharge dusty plasma in the background of Ar. gas. The experiments have been carried out in a versatile table-top Dusty Plasma Experimental (DPEx) device which is described in detail elsewhere \cite{surabhi2015}. The flow of the dust fluid is induced by suddenly reducing the neutral gas flow rate (by creating a density gradient). An isolated copper wire has been used to create an electrostatic potential hill that acts as an obstruction to the flow of dust fluid. As a result, the particles are accumulated at the base of the potential hill created by the wire which later leads to the formation of shock fronts.
  A detailed characterization of the propagation characteristics of these structures is made by changing the height of the potential hill and the background gas flow rate. 
   Our experimental results are compared to model calculations carried out on a KdV-Burger type equation that is derived from the Generalized Hydrodynamic model equations and found to be in good agreement with the theoretical predictions. \par
The paper is organized as follows.  In the next section (Sec.~\ref{sec:setup}), we describe our experimental set-up. In Sec.~\ref{sec:results}, we present our experimental results on the shock excitation and a detailed  characterization of their dynamics. A theoretical description based on the model KdV-Burger equation is presented in sec.~\ref{sec:theory} and a comparison is made with the experimental findings. Sec.~\ref{sec:conclusion} provides a summary of our work and some concluding discussion and observations.
\section{Experimental Set-up}\label{sec:setup}
\begin{figure}[ht]
\includegraphics[scale=0.48]{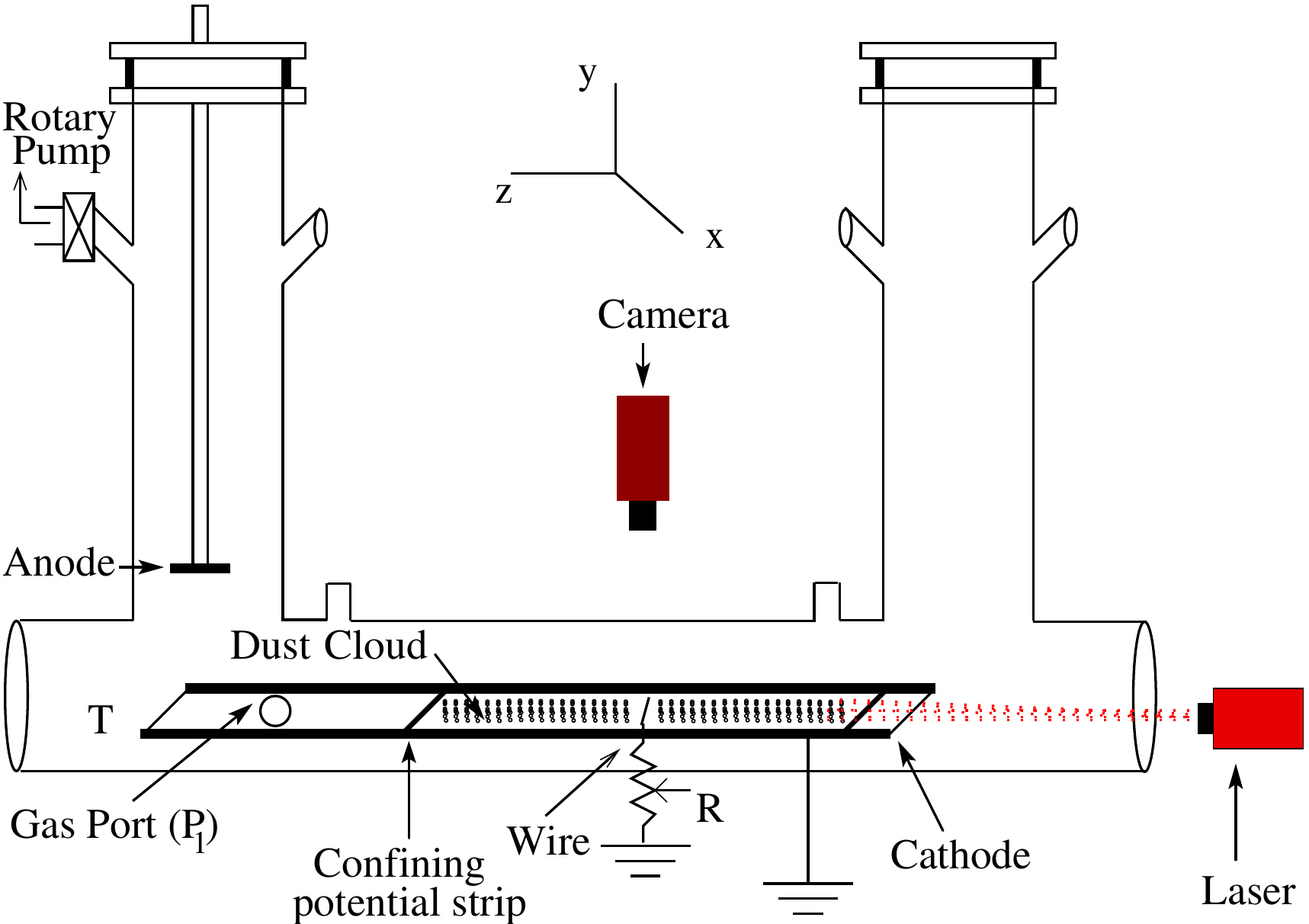}
\caption{\label{fig:setup} A schematic of the Dusty Plasma Experimental (DPEx) Device. T: grounded cathode tray, R: variable resistance to change the height of the potential hill.}
\end{figure}

Our experiments have been performed in a Dusty Plasma Experimental (DPEx) device shown schematically in Fig.~\ref{fig:setup}. For a detailed description of the experimental device the reader is referred to Ref. \cite{surabhi2015}. The vacuum chamber is a $\Pi$-shaped glass tube with several axial and radial ports for different purposes. A stainless steel (SS) disc electrode of 3 cm diameter (insulated with ceramic from the back side), located at the left arm of the $\Pi$-tube, serves as an anode whereas a grounded SS tray of 40 cm $\times$ 6.1 cm $\times$ 0.2 cm is placed on the connecting tube and used as a cathode. A couple of SS strips are placed approximately $25~cms$ apart on the cathode to confine the dust particles in the axial direction by means of their sheath electric fields. A copper wire  of 1 mm diameter, insulated by ceramic beads, is mounted in between these two strips at a height of $1~cm$ from the cathode. Normally the wire is kept at the floating potential but there is a provision to make the wire acquire a ground potential. In fact, we are able to adjust the potential of the wire to different potential levels by connecting a variable resistance ranging from 10 k$\Omega$ to 10 M$\Omega$ as shown in Fig.~\ref{fig:setup}. \par 
To create the dusty plasma, micron sized kaolin particles (with a size dispersion ranging from $2$ to $6$ microns) are sprinkled on the cathode tray before closing the experimental device.
 The experimental chamber is  evacuated to a base pressure of $10^{-3}$ mbar by a rotary pump.
Argon gas is then flushed into the chamber through a mass flow controller (which can be used to control the flow rate of neutral gas very precisely) and then pumped down to the base pressure. This process is repeated  several times to avoid any kind of impurity in the plasma. Finally the background pressure is set to a working pressure of $P=0.110-0.120$ mbar by adjusting the pumping speed and the gas flow rate. In this condition, the gate valve (connected on the mouth of the pump) is approximately opened to 20\% and the mass flow controller is opened to $10\%$ (corresponding to flow rate of 27.5 m$l_s$/min) to maintain a constant pressure. The direct current (DC) glow discharge plasma is then produced in between the anode and the grounded cathode by applying a voltage, $V_a = 400$~V. The applied voltage is then reduced to $360-370$~V at a discharge current of $I_p \sim 4$~mA.  The plasma parameters such as ion density ($n_i$), electron temperature ($T_e$) and plasma/floating potentials are initially measured using single Langmuir and emissive probes in the absence of dust particles. The typical experimentally measured values of these plasma parameters are $n_{i} \approx 10^{15}~m^{-3}$, $T_{e} \approx 4-5$~eV.  The ions are assumed to be at room temperature i.e., $T_{i} \approx 0.03$~eV\cite{thompsion}. 
For the given discharge voltage of $370$~V, a dense dust cloud is found to levitate near the cathode sheath boundary. The levitated dust particles are illuminated by a red diode laser and the scattered light from the dust cloud is captured by a fast CCD camera (130 fps) with 1000 pixels $\times$ 225 pixel resolution. From the analysis of the video images the dust density ($n_{d0}$) and temperature ($T_d$) are estimated to be $n_{d} \approx 10^{11}~m^{-3}$ and $0.6-1.5~eV$\cite{surabhi2015}. The average dust mass is $m_{d} \approx 10^{-14}-10^{-13}~kg $ and the average charge on a dust particle (inferred from the plasma parameters and the particle size) is approximately $Q_{d} \approx 10^{4} e$. The electron density is obtained from the quasi-neutrality condition by taking account of the dust contribution. Based on these values the typical magnitude of the linear phase velocity of a dust acoustic wave (DAW) for our experimental conditions turns out to be $v_{ph} \sim 4-5$~cm/sec. Using these plasma and dusty plasma parameters, the screened Coulomb coupling parameter, $\Gamma=\frac{Q_d^2}{4\pi\epsilon_0 d KT_d}exp(-\kappa)$ \cite{jyoti2002}, is estimated as $ \sim 109$ for a screening parameter of $\kappa = d/\lambda_i \sim 3.9$, where $d\sim168$ $\mu$m is the inter-particle distance, $T_d=1.2$~eV and $\lambda_i \sim 43$  $\mu$m is the ion Debye screening length.\par
To start with we adjust the pumping speed and the gas flow rate in a precise manner to achieve a stationary dust cloud in the region to the right of the wire mounted over the cathode. This equilibrium state is achieved at $\sim 10\%$ opening of the mass flow controller and $\sim $ 20$\%$ opening of the gate valve. Now when we change the gas flow rate near the gas feeding port ($P_1$) as shown in Fig.~\ref{fig:setup}, the particles are found to move either towards the direction of the pump (in case of decreasing flow rate) or away from the pump  (in case of increasing flow rate). A detailed study on the flow generation techniques has been reported elsewhere \cite{surabhi2015}.  In the present set of experiments, the flow of the dust fluid over the wire is generated by reducing the gas flow rate in steps of 1$\%$ which corresponds to 2.75 $ml_s$/min. \\
\section{Excitation of Shock Waves and its characterization}\label{sec:results}
\begin{figure}[htb]
\centering
\includegraphics[width=0.45\textwidth]{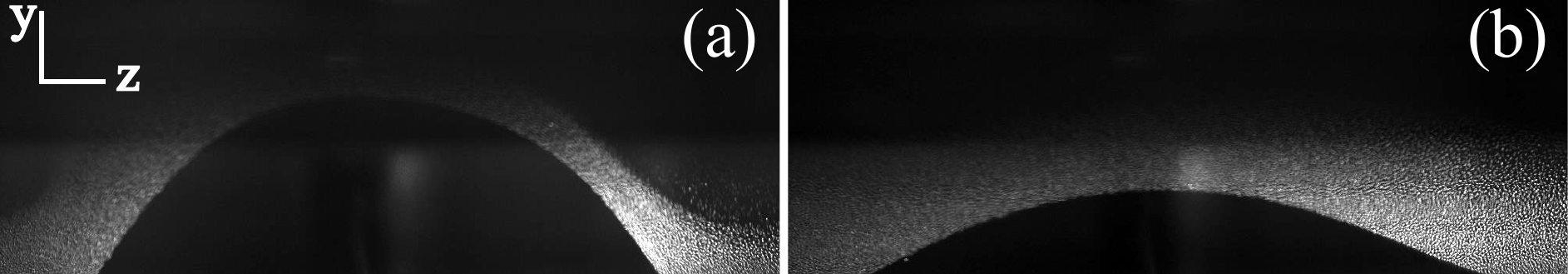}
\caption{Potential profile for the a) grounded wire b) floating wire. The height of the potential hill decreases considerably when the wire is switched to floating potential from grounded potential. } 
\label{fig:potentialprofile}
\end{figure}

As discussed in the previous section, the trapped stationary dust cloud is allowed to flow from right to left by reducing the neutral gas flow rate suddenly from its equilibrium condition for a time duration less than a second. When this change of background gas flow rate is very small (e.g. when the opening of mass flow controller changes from 10$\%$ to 5$\%$ or less), it is found that the dust fluid simply flows over the potential hill created by the floating wire with the grounded wire merely creating a hindrance in the path of the particle flow. Fig.~\ref{fig:potentialprofile} shows snap shots of particle flow in the Y-Z plane when the wire is grounded (Fig.~\ref{fig:potentialprofile}(a)) and when the wire is allowed to be at the floating potential (Fig.~\ref{fig:potentialprofile}(b)). The profiles of the particle trajectories in these two different situations are seen to closely follow the potential profiles created by the wire. It can be seen that the height of the potential hill is reduced significantly when the wire potential is switched from the ground potential to a floating potential.\par
\begin{figure}[htb]
\centering
\includegraphics[width=0.45\textwidth]{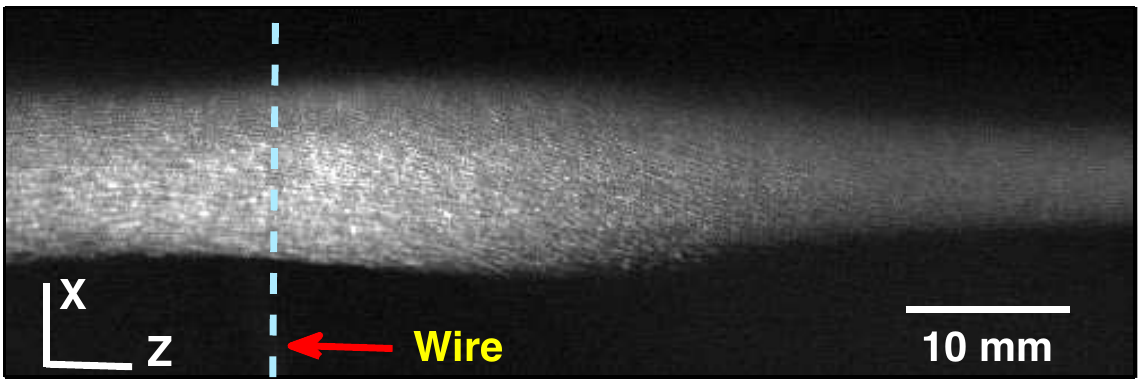}
\caption{Dust fluid flow over the floating wire when the flow rate difference is 8$\%$. Dashed line represents the location of the wire.} 
\label{fig:onlyflow}
\end{figure}
Fig.~\ref{fig:onlyflow} depicts the flow of dust fluid over the floating wire in X-Z plane when the background gas flow rate is changed from 10$\%$ to 2$\%$ (27.5 ml$_s$/min to 5.50 ml$_s$/min) in a time less than a second. The dashed line represents the location of the wire. Since the wire is at the floating potential, which is approximately equal to the surface potential of the dust, the dust fluid does not feel the presence of the wire and hence it simply exhibits streamline flow over the wire in the entire range of flow rate change.  
\par 
\begin{figure}%
    \centering
    \subfloat{{\includegraphics[scale=0.7]{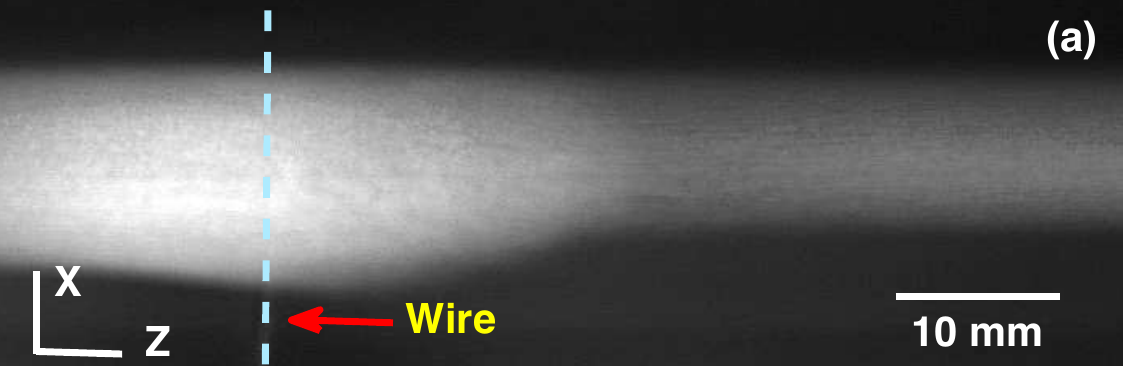}}}%
  \vspace*{-0.13in}
    \subfloat{{\includegraphics[scale=0.7]{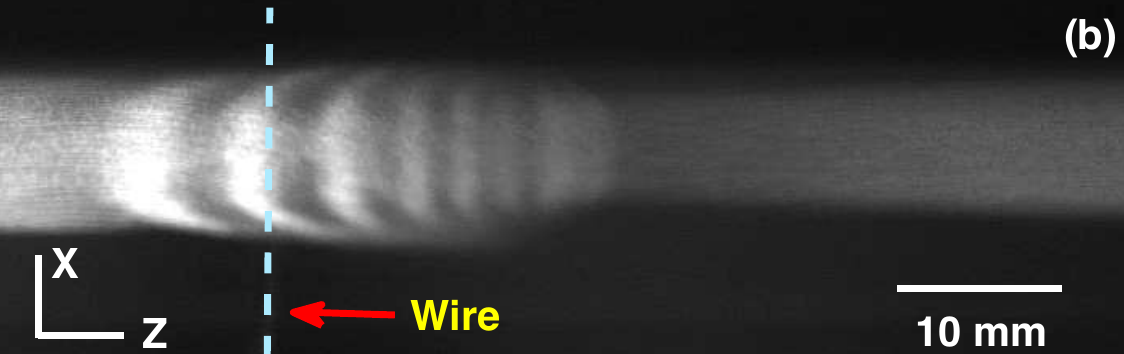} }}%
    \caption{Typical image of (a) Transition stage and  (b) Oscillatory shock wave fronts.}%
    \label{fig:structure}
\end{figure}
To excite the dust acoustic shock waves \cite{merlino_1999,pakzad_2011}, the wire is kept at the ground potential. The sheath around the wire acts as a potential barrier (or hill) to the flowing dust fluid. The dust fluid is then forced to flow over the potential hill by reducing the gas flow rate suddenly from 10$\%$ (27.75 m$l_s$/min) to below 5$\%$ (13.75 m$l_s$/min) at point $P_1$.  Due to the sudden decrease of neutral density at point $P_1$, the neutrals rush towards this point from all directions to attain a homogeneous equilibrium neutral density. As a result, the dust particles (carried by the neutrals) are also found to flow towards the pump with a supersonic velocity\cite{surabhi2015}. The potential hill created by the wire opposes the free motion of the particles, resulting in a compression of the dust cloud near its base. Consequently it leads to a sudden density jump near the wire which subsequently expands and form a shock that is found to propagate in the direction of the fluid flow. The transient stage of density accumulation is shown in fig.~\ref{fig:structure}(a). Afterwords with the evolution of time, this transition is transformed into a dispersive DASW  by the creation of compression and rarefaction in the dust density near the wire.

 Fig.~\ref{fig:structure}(b) displays a typical image of the curved dispersive shock  fronts. 
  The range of discharge parameters where the waves get triggered is $P = 0.10-0.12$ mbar, $V_a \sim 360-370$ V and $I_p \sim 4$ mA.  It is also worth mentioning that there exists a threshold (13.75 m$l_s$/min.) in the difference of gas flow rate  below which the DASWs do not get triggered, because in that case, the fluid flow remains subsonic.\par
\begin{figure}[htb]
\centering
\includegraphics[width=0.35\textwidth]{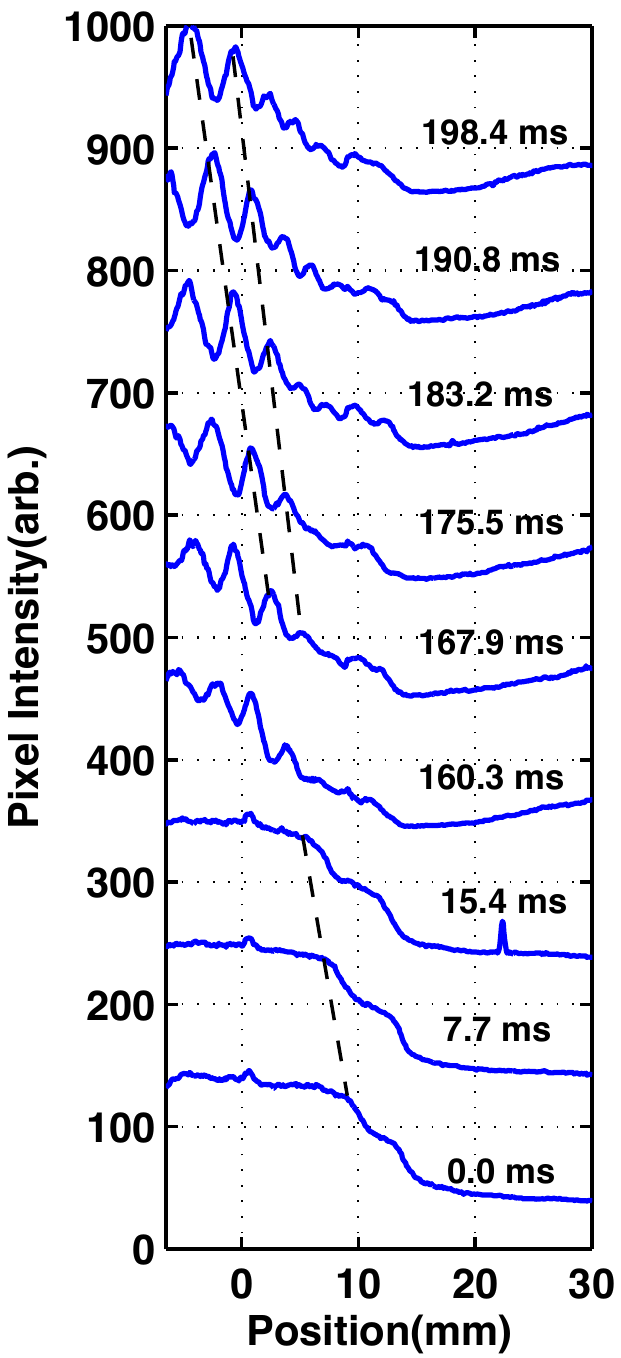}
\caption{(Color online) Time evolution of oscillatory dust acoustic shock fronts. The \lq 0' position in the plot corresponds to the position of the  wire connected to the ground. Top most plot (t=198.4 ms) corresponds to the Fig.~\ref{fig:structure}(b).} 
\label{fig:evolution}
\end{figure}
Fig.~\ref{fig:evolution} shows the time evolution of a dispersive DASW. The \lq 0' position represents the location of the wire. As discussed above, it can be seen from the figure that due to sudden jump of dust density near the wire (as seen from the first three curves of Fig.~\ref{fig:evolution}), a shock wave is generated and later (after $\sim 70$ ms, although in the figure the time evolution is shown after 160 ms when the small fluctuations in the dust density fully converts into the shock fronts) a small oscillation develops near the base of the potential hill which grows with time and evolves into an oscillatory shock front. As a result, the pulse grows in amplitude as it propagates towards the wire and after crossing the wire the amplitude of the pulse decreases whereas the width increases. Depending upon the discharge condition and the change of gas flow rate the dispersive dust acoustic shock propagate with a velocity in the range of $10-20$ cm/sec. \par
\begin{figure}
\centering
\includegraphics[width=0.35\textwidth]{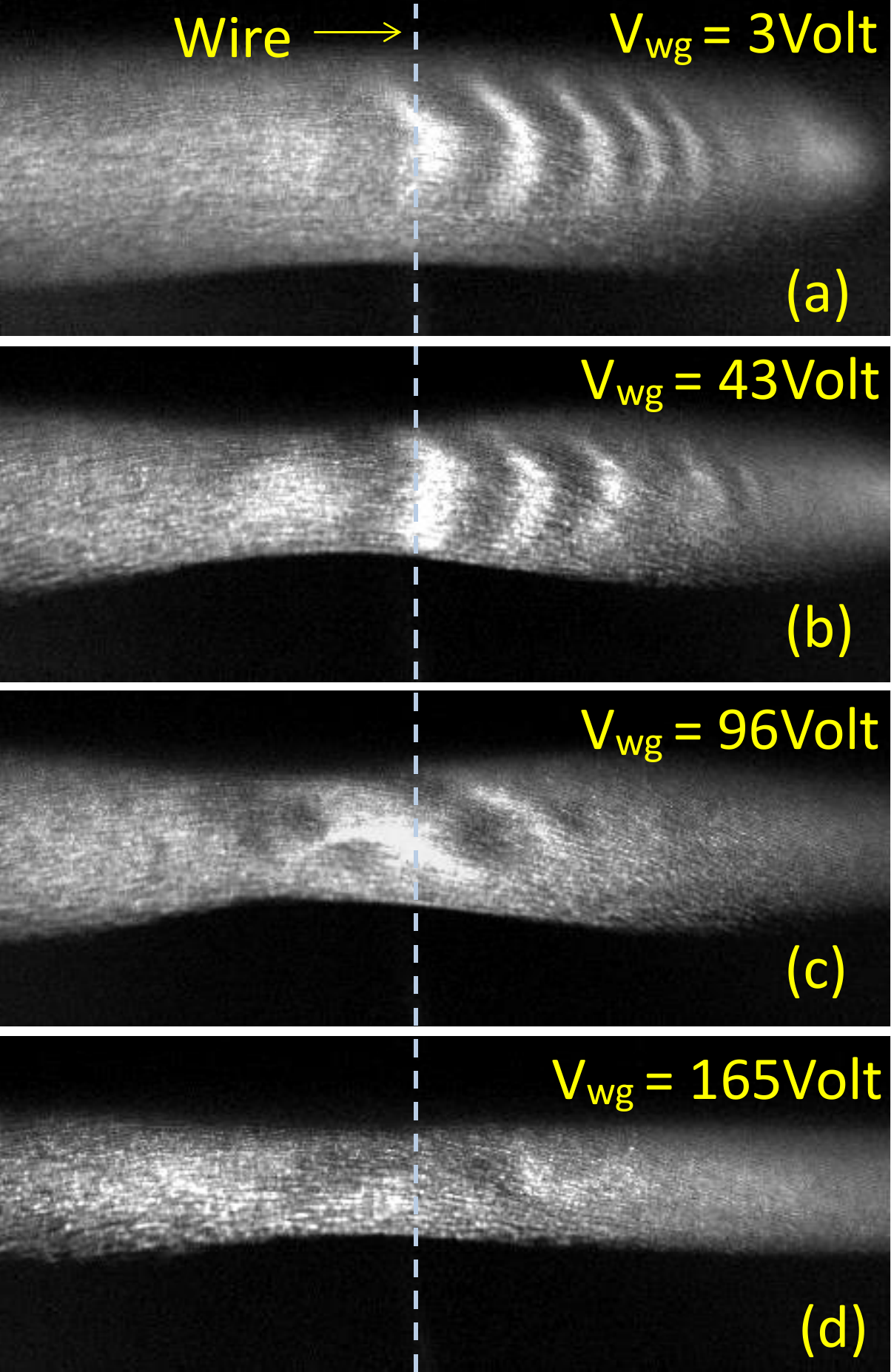}
\caption{(Color online) Images of shock waves for different values of potential hill for a) 3 Volt, b) 43 Volt, c) 96 Volt, d) 165 Volt. $V_{wg}$ is the potential of the wire with respect to the grounded cathode. The dashed line is representing the position of the wire which acts as potential hill for the flow of dust fluid.} 
\label{fig:flowresistance}
\end{figure}
In addition to the threshold in the flow rate, there is also a threshold in the height of the potential hill above which shock waves get triggered.  Fig.~\ref{fig:flowresistance} shows the flow of the dust fluid over the wire for different heights of the potential hill. A resistance bank of variable resistance from 10 k$\Omega$ to 10 M$\Omega$, connected between the wire and the ground, is used to change the height of the potential hill. By drawing current through different combinations of resistances one can change the height of the potential hill ($V_{wg}$). As is well known, dust particles at equilibrium do not draw a net current ($I_e=I_i$), which implies that they have nearly the same potential as that of the floating wire.  Therefore, when the dust fluid flows over the floating ($V_f=V_{wp}\sim 250$ V) wire it barely feels the presence of a potential hill  whereas it experiences the maximum obstruction from the potential hill when it flows over the grounded wire ($V_{wp}=0$ V). Fig.~\ref{fig:flowresistance}(a) shows the excitation of shock waves when the potential of the wire is 3 V which is approximately the value of the ground potential. If the  wire potential with respect to ground keeps on increasing towards the floating potential the dust fluid faces a smaller and smaller hindrance in its path of flow. Fig.~\ref{fig:flowresistance}(b) shows that the number of shock fronts decreases (from five to three) with the decrease of the height of the potential hill. If we increase the wire potential further with respect to ground (hence decrease the height of the potential hill), we find a threshold value beyond which we could not excite DASWs. In our experiments the threshold value is found to be $\sim 50$ V. If the wire potential is increased further (see Fig.~\ref{fig:flowresistance}(c)), the structures becomes unstable and break up into turbulence. In Fig.~\ref{fig:flowresistance}(d), the potential of the wire is 165 V, which is quite far from the floating potential but sufficiently large such that potential hill over the wire is  very small and therefore incapable of exciting shock waves. Consequently we observe an almost streamline motion of the dust particles over the wire. \par
\begin{figure}[!b]
\centering
\includegraphics[width=0.45\textwidth]{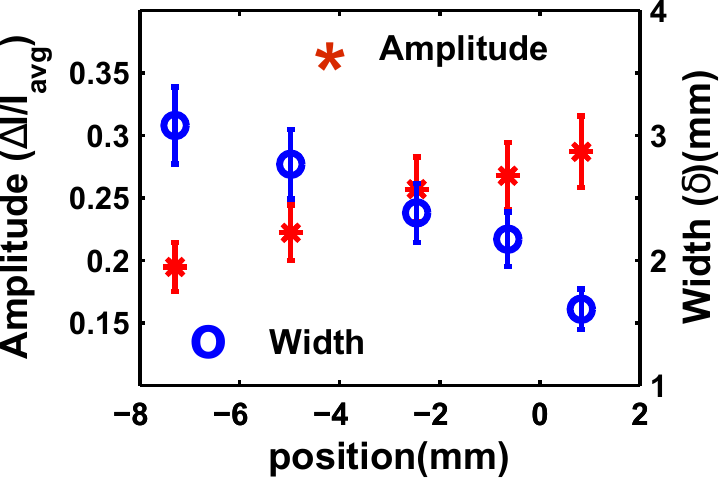}
\caption{(Color online) Variation of amplitude ($*$) and shock thickness ($o$) of a leading oscillatory shock front with position. In this set of experiments the change of flow rate is 19.25 ml/min.} 
\label{fig:ampwidth}
\end{figure}
The characteristics of the observed shock waves have been studied by measuring their amplitudes ($\Delta n_d/n_{d0}$) as well as the thicknesses ($\delta_w$) of the leading pulses of dispersive shock fronts when the small fluctuations in the dust density are completely transformed into oscillatory shock fronts. These shock parameters are obtained by following the technique used by Heinrich \textit{et al.} \cite{heinrich2009} and Annibaldi \textit{et al.}\cite{annibaldi2007}. The amplitude which is also proportional to $\Delta I/I_{avg}$ is calculated from the measured pixel intensities $I$, where $\Delta I$ ($= I_{max}-I_{avg}$) is the difference between the maximum intensity ($I_{max}$) of the leading shock front from its average value ($I_{avg}$). The shock thickness ($\delta_w$) is defined as the difference between the steep edge point and the peak point of a crest. Fig.~\ref{fig:ampwidth} shows the variation of amplitude (indicated by \lq$*$') and shock thickness (indicated by \lq o') of oscillatory DASWs with position.
The \lq0' level in the figure corresponds to the position of the wire which is taken as reference point for the propagation of dispersive dust acoustic shock waves (DASW). It can be seen in the figure that the shock amplitude  follows approximately a linear decay whereas the thickness  increases as the shock front propagates away from the wire. The amplitude fall off is a possible evidence of dissipation within the shock which is similar to that observed in the gas dynamic shocks and can be attributed to the effects of viscosity\cite{heinrich2009,samsonov2003}.
\par
\section{Comparison of experimental observations with theoretical modeling results}\label{sec:theory}
In this section, the time evolution of large amplitude dust acoustic shock waves obtained in experiments is compared with analytical solutions of a model Korteweg-de Vries-Burgers (K-dV-Burgers) type equation\cite{shukla2001} that is relevant for strongly coupled dusty plasmas. To derive such an equation, we consider an unmagnetized strongly coupled dusty plasma whose constituents are electrons, ions, and negatively charged massive dust grains. The charge neutrality condition at equilibrium for such a system is given by $n_{i0}=Z_dn_{d0}+n_{e0}$, where $n_{i0}$, $n_{d0}$ and $n_{e0}$ are the unperturbed ion, dust, and electron number densities, respectively, and $Z_d$ is the number of charges residing on the dust grain surface. The lighter ion and electron species can be described by Boltzmann distributions at temperatures $T_i$ and $T_e$ respectively and are coupled weakly to each other due to their higher tempratures and smaller electric charges. The dust grains of mass $m_d$, on the other hand, are strongly coupled to each other because of their lower temperature ($T_d$) and larger electric charge ($Q_d=Z_de$). Considering a weakly collisional dusty plasma ($\nu_{dn}\tau_m\rightarrow0$), the dynamics of the nonlinear DAWs in such a strongly coupled dusty plasma can be well described by the generalized hydrodynamic (GH) model\cite{kaw1998}, whose governing equations are given by,
\begin{eqnarray} 
\frac{\partial  n_d}{\partial  t}+\frac{\partial}{\partial x}(n_d u_d)& =& 0,\label{eqn:continuity}\\
D_t u_d + \nu_{dn} u_d - \frac{\partial \phi}{\partial x}&=&\frac{\eta_l}{n_d}\frac{\partial^2u_d}{\partial x^2},
\\\label{eqn:momentum}
\frac{\partial^ 2\phi}{\partial x^2} = n_d+\alpha_e e^{\sigma \phi}-&\alpha_i e^{- \phi}&.\label{eqn:pos}
\end{eqnarray}
where $n_d$  is the instantaneous  dust density normalized by its equilibrium value $n_{d0}$, $u_d$  is the  dust fluid velocity normalized by the dust acoustic speed $C_d=(Z_dT_i/m_d)^{1/2}$, $\phi$ is the electrostatic wave potential normalized by $T_i/e$. The constants quantities are defined as,  $\alpha_e=n_{e0}/Z_d n_{d0}$, $\alpha_i=n_{i0}/Z_d n_{d0}$ and $\sigma=T_i/T_e$. The time and space variables are in units of the dust plasma period $\tau_d=(m_d/4\pi n_{d0} Z_d^2 e^2)^{1/2}$ and the Debye length $\lambda_{Dd}=(T_i/4\pi Z_d n_{d0} e^2)^{1/2}$, respectively and $D_t=\partial/\partial t + u_d \partial/\partial x$. Furthermore, $\nu_{dn}$ is the dust-neutral collision frequency normalized by the dust plasma frequency $\tau_d^{-1}$ and $\tau_m$ is the relaxation time. The combined viscosity can be expressed as, $\eta_l=(\tau_d/m_d n_{d0}\lambda_{Dd}^2)[\eta_t + (4/3)\zeta_t]$,\cite{shukla2001} where $\eta_t$ and $\zeta_t$ represent shear and bulk viscosities.\par
To derive the K-dV-Burgers equation, we have used the reductive perturbation technique with the stretched coordinates 
\begin{eqnarray} 
\xi&=&\epsilon^{1/2}(x-v_0 t),\label{eqn:strech1}\\ 
\tau&=&\epsilon^{3/2} t.\label{eqn:strech2}
\end{eqnarray}
where $\epsilon$ is the smallness parameter measuring the weakness of the amplitude or dispersion and $v_0$ is the wave phase velocity (normalized by $C_d$). In the expansion of variables $n_d, u_d$  and $\phi$, we include the equilibrium flow velocity (i.e. the terminal velocity \cite{surabhi2015}) of fluid  $u_{d0}$ at the equilibrium condition in accordance with our experiments. The variables can be expanded as
\begin{eqnarray}
&n_d& = 1+ \epsilon n_d^{(1)}+\epsilon^ 2 n_d^{(2)}+...,\label{eqn:diff asymnd}\\
&u_d &= u_{d0}+\epsilon u_d^{(1)}+\epsilon^ 2 u_d^{(2)}+...,\label{eqn:asymvd}\\
&\phi& =\epsilon \phi^{(1)}+\epsilon^2 \phi^{(2)}+...,\label{eqn:asymphi}\\
&\eta_l& = \epsilon^{1/2}\eta_0.\label{eqn:eta_expansion}
\end{eqnarray}
where $\eta_0$ is the kinematic viscosity \cite{shukla2001}.\\
{\color{black}Substituting Eq.~(\ref{eqn:strech1})-(\ref{eqn:eta_expansion}) into Eq.~(\ref{eqn:continuity})-(\ref{eqn:pos}), we obtain from the equation of the lowest order in $\epsilon$,
\begin{eqnarray}
 &u_d^{(1)}&=\frac{-\phi^{(1)}}{(v_0-u_{d0})}\\
 &n_d^{(1)}&=\frac{-\phi^{(1)}}{(v_0-u_{d0})^2}\\
 &v_0&=u_{d0}+\frac{1}{(\sigma\alpha_e+\alpha_i)^{1/2}}. 
\end{eqnarray}}
 Using the expressions obtained from the first order calculation we get the following equation from the next lowest order of $\epsilon$:
\begin{equation}
 A^{-1}\frac{\partial \phi^{(1)}}{\partial \tau} + \phi^{(1)}\frac{\partial \phi^{(1)}}{\partial \xi} + \beta \frac{\partial^3 \phi^{(1)}}{\partial \xi^3}=\mu \frac{\partial^2 \phi^{(1)}}{\partial \xi^2}
 \label{eqn:burger}
\end{equation}
where the constants are give by the following expressions,
\begin{eqnarray}
\delta&=&-\frac{3}{(v_0-u_{d0})^4}-\sigma^2\alpha_e + \alpha_i  \nonumber\\
A&=&\delta(v_0-u_{d0})^3 /2 \nonumber \\
\mu&=&\eta_0/[\delta(v_0-u_{d0})^3] \nonumber\\
\beta&=&1/\delta \nonumber
\end{eqnarray}
To obtain a stationary solution of the K-dV Burgers equation we further transform to a frame defined by $\zeta=\xi-U_0\tau$, and $\tau=\tau$. Eqn.(\ref{eqn:burger}) can then be integrated once and converted to,
\begin{equation}
\beta\frac{\partial^2 \phi^{(1)}}{\partial\zeta^2}-\mu\frac{\partial\phi^{(1)}}{\partial\zeta}+\frac{1}{2}[\phi^{(1)}]^2-\frac{U_0}{A}\phi^{(1)}=0 \label{eqn:burgersolution}
\end{equation}
A monotonic shock wave is formed when the dissipation term (second term of Eqn.(\ref{eqn:burgersolution})) dominates over the dispersion term (first term of Eqn.(\ref{eqn:burgersolution})). In this case the solution of Eq.~(\ref{eqn:burgersolution}) can be expressed as : 
\begin{equation}
\phi^{(1)}=\frac{U_0}{A}\left[1-\text{tanh}\left(\frac{U_0}{2A\mu}(\xi-U_0\tau)\right)\right].\label{eqn:monotonic}
\end{equation}
where, $U_0$, $\phi_0=U_0/A$ and $\Delta=A\mu/U_0$ are the speed, height and thickness of the shock fronts, respectively. It is worth mentioning that for a given $\mu$, if the amplitude of the shock front decreases the shock thickness increases and vice versa. It is also to be noted that if $\mu$ is extremely small, the shock will have an oscillatory profile in which the first few oscillation at the wave front will be close to a solitonic form \cite{hamid2009}. There exists a critical value of dissipation coefficient $\mu$ that determines whether a  monotonic or an oscillatory shock solution \cite{shukla2001} is formed. This critical value of $\mu$, as determined from the theoretical model, is $\mu_c=(4\beta U_0/A)^{1/2}$. The shock wave has a monotonic profile for $\mu^2>\mu_c^2$ and an oscillatory profile for $\mu^2<\mu_c^2$. For $\mu^2 < \mu_c^2$, the oscillatory solution is given as 
\begin{equation}
\phi^{(1)}=\varphi+C\text{exp}\left(\frac{Z\mu}{2\beta}\right)\text{cos}\left(Z\sqrt{\frac{U_0}{A\beta}}\right) \label{eqn:oscillatory}
\end{equation}
where, $Z=\xi-U_0\tau$ and $C$ is a constant. It should be noted that for a weakly collisional dusty plasma ($\nu_{dn}\tau_m\rightarrow$0), we have $\delta<0$, which corresponds to A$<0$, $\mu<0$, and $\beta<0$.\par
For the completeness of this study, we now make an attempt to compare qualitatively the results obtained in experiments with the analytical results we get by solving the the KdV-Burgers equation. For this comparison, the experimentally measured plasma and dusty plasma parameters are used to estimate the constants associated with KdV-Burgers equation.  For the given values of discharge and plasma parameters in our experiments, the coefficients of KdV-Burgers equation come out to be $\delta=-1.36$, $A = -0.85$, $\beta=-0.73$ and $\mu=-0.12$. The kinematic viscosity at $\Gamma \sim 109$ is taken as $\eta_0 =0.2$\cite{ichimaru1987} as in the experiments of Nakamura {\it{et al.}} \cite{nakamura2002}. This highly viscous dusty plasma fluid (due to the strong coupling between the particles) provides necessary dissipation in the medium to form stable shock fronts. {\color{black} As the magnitude of fluid velocity determines the strength of the perturbation in the dust medium, hence this flow also plays an important role to determine the shape and size of the shock structure. The analytical solutions  (Eq. \ref{eqn:oscillatory}) of KdV-Burgers equation for $u_{d0}=8$~cm/sec at $\eta_0 =0.2$ is plotted in Fig.~\ref{fig:burger}(a). It shows the oscillatory shock fronts appear when there is a sudden jump of perturbed dust density. Similar trend of shock fronts have also been observed in our experiments (as shown in Fig.~\ref{fig:burger}(b)) when the background intensity is subtracted from raw data.  In addition, it is also found that when the fluid flow is increased further (not shown in the figure), the velocity as well as the amplitude of the shock waves increase whereas the width decreases which essentially shows a clear signature of nonlinearity in the structure.  This amplitude-width relationship is also revealed from experimental observation which is shown in the Fig.~\ref{fig:ampwidth}. \par
\begin{figure}[htb]
\centering
\includegraphics[width=0.45\textwidth]{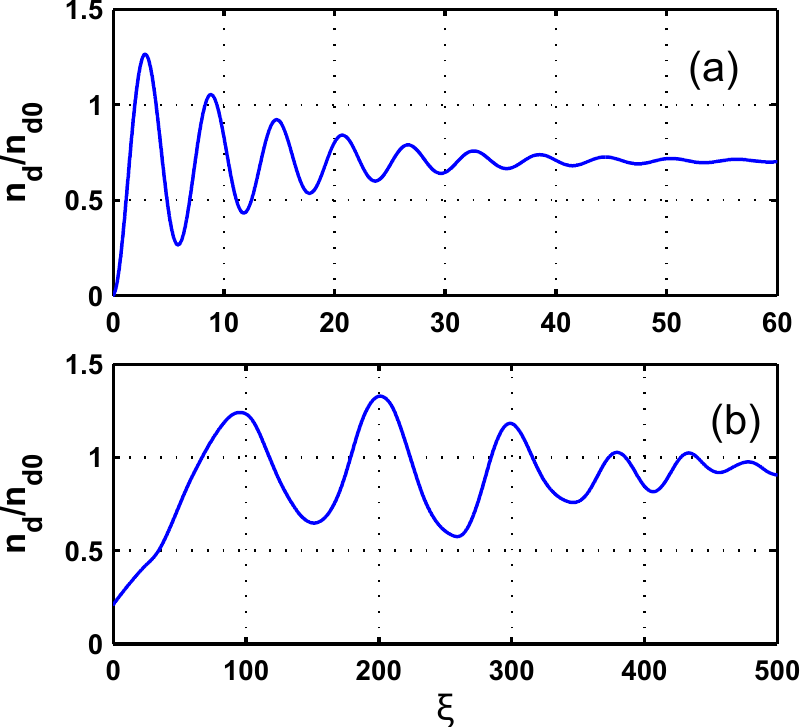}
\caption{(Color online) Nature of shock waves find a) theoretically and b) experimentally. The analytical results are plotted for fluid velocity 8 cm/sec and $\eta=0.2$.} 
\label{fig:burger}
\end{figure}
The above agreement in the nature of the experimental observations and the theoretical results suggests  that the Generalized-Hydrodynamic model\cite{kaw1998} is able to capture the essential underlying dynamics governing the formation of these shock structures both in terms of the existence  and observed trend in the variation of the amplitude and thicknesses of the shock structures. The model however falls short of providing  an accurate quantitative measure of the shock thickness - the experimental values are about 5 to 10 times higher than the theoretical values. 
This mismatch indicates that some additional dissipation processes other than the dissipation due to strong correlation between the particles could be playing a role in the experimental situation and needs to be incorporated in the model. Since in our experiments we excite the shock waves by introducing a strong neutral gas flow, this additional dissipation could come from dust-neutral collisions which have been neglected in the model.  
The shock thickness in such a case can be approximately equal to the dust-neutral collision mean free path \cite{heinrich2009,samsonov2003} and can be defined as $\delta_w = \lambda_{dn} = u_{do}/\nu_{dn}$. The dust-neutral collision frequency ($\nu_{dn}$) can be obtained from the standard relation for the  Epstein\cite{epstein1924} drag. For a pressure of 0.12 mbar and dust radius $a=2 ~\mu m$, $\nu_{dn}$ turns out for for our experimental conditions to be $\sim18 s^{-1}$. The typical shock thicknesses, based on the above parameters, are then found to be $\delta_w \sim 3.6$~mm for $u_{do} = 6.5$~cm/sec
which is quite close to the experimental values shown in Fig.~\ref{fig:evolution} and Fig. \ref{fig:ampwidth}. 
\section{Conclusion}
\label{sec:conclusion}
{\color{black} In conclusion, we have experimentally demonstrated the formation of dispersive shock waves in DPEx device by generating a flow of dust fluid over a grounded wire. The flow of dust fluid is initiated by sudden reduction of mass flow rate of neutral gas. This instantaneous change of flow rate creates a dust density jump near the potential hill, created by the grounded wire. Eventually this jump of density propagates in the form of oscillatory shock fronts. The propagation characteristics of these oscillatory shock structures have been characterized thoroughly by varying the strength of the potential hill and the gas flow rate at a constant discharge parameters. The amplitude of the shock fronts decays whereas the shock thickness increases when it propagates away from the wire. The nature of shock waves become turbulent through a intermediate state when the height of the potential hill reduces to a small value (i.e., when the potential of the wire changes from being grounded to being floating). It is also observed that there exists a threshold of the height of the potential hill below which the shock waves could not be excited.
A Generalized Hydrodynamic equation has been used to model the experimental observation of shock formation by accounting for the effect of dust-dust interaction in the form of viscosity and to derive a KdV-Burgers equation including the effect of flow velocity. This model explains qualitatively the formation of  shock waves and the changes in its nature with the change of flow velocity. Moreover, the model also validates the variation of experimentally obtained amplitude and width of shock fronts. An effect of dust-neutral collision is also introduced to delineate the exact dissipation mechanism. Our experimental observation of shock formation in a flowing dusty plasma besides providing insights into the nonlinear dynamics of complex plasmas, can also be of potential value in understanding shock phenomena in space plasmas such as during solar wind interactions with the earth as well as in astrophysical scenarios where shock waves are a common occurrence in galactic dust clouds during the early stages of star formation.}
\begin{acknowledgments} 
We are thankful to Dr. G. Ravi for his valuable discussion and suggestions on our experiment. One of the author,  A.S. gratefully acknowledges the support provided for this
research by Grant No. FA2386-13-1-4077 AOARD 134077.
\end{acknowledgments} 


\begin{thebibliography}{1}
\bibitem{Ikezi}
H. Ikezi, Phys. Fluids {\bf{29}}, 1764 (1986).
\bibitem{thomas_1994}
H. Thomas, G. E. Morfill, and V. Demmel, Phys. Rev. Lett. {\bf{73}}, 652-655 (1994).
\bibitem{ma}
P. K. Shukla and V. P. Silin, Phys. Scr. {\bf{45}}, 508 (1992).
\bibitem{rao}
N. N. Rao, P. K. Shukla, and M. Y. Yu, Planet. Space Sci. {\bf{38}},
543 (1990).
\bibitem{melands}
F. Melands\O, Phys.Plasmas {\bf{3}}, 3890 (1996).
\bibitem{N}
N. N. Rao, Phys.Plasmas {\bf{6}}, 4414 (1999).
\bibitem{goree}
D. Samsonov and J. Goree, Phys.Rev.E {\bf{59}}, 1047 (1999).
\bibitem{law}
D. A. Law, W. H. Steel, B. M. Annaratone, and J. E. Allen, Phys.
Rev. Lett. {\bf{80}}, 4189 (1998).
\bibitem{bailung}
Y. Nakamura and H. Bailung,Phys. Rev. Lett {\bf{83}}, 8 (1999).
\bibitem{Andersen}
H. K. Andersen, N. D'Angelo, P. Michelsen, and P. Nielsen. Phys. Rev. Lett. {\bf{19}}, 149 (1967).
\bibitem{shuk}
N. Rao and P. Shukla, Planet. Space Sci. {\bf{42}}, 221 (1994).
\bibitem{kour}
I. Kourakis and P. K. Shukla, Eur. Phys. J. D {\bf{29}}, 247 (2004).
\bibitem{morf}
G. E. Morfill and A. V. Ivlev, Rev. Mod. Phys. {\bf{81}}, 1353 (2009).
\bibitem{versheest}
F. Verheest, Space Sci. Rev. {\bf{77}}, 267 (1996).
\bibitem{mamun}
A. A. Mamun and P. K. Shukla, Phys. Scr.,  {\bf{T98}}, 107 (2002).
\bibitem{samsonov}
D. Samsonov, A. V. Ivlev, R. A. Quinn, G. Morfill, and S. Zhdanov, Phys. Rev. Lett. {\bf{88}}, 095004 (2002).
\bibitem{Pintu}
P. Bandyopadhyay, G Prasad, A. Sen, and P.K. Kaw, Phys. Rev. Lett. {\bf{101}}, 065006 (2008).
\bibitem{heidemann}
R. Heidemann, S. Zhdanov, R. Sutterlin, H. M. Thomas, and G. E. Morfill, 
Phys. Rev. Lett  {\bf{102}}, 135002 (2009).
\bibitem{j.bond.1965}
J. Bond, K. Watson, and J. Welch, Atomic Theory of Gas Dynamics
(Addison-Wesley, Reading, MA, 1965).
\bibitem{dimon 1999}
S. Ho"rluck and P. Dimon, Phys. Rev. E {\bf{60}}, 671 (1999).
\bibitem{R. Graham1993}
R. Graham, Solids under High-Pressure Shock Compression
(Springer-Verlag, New York, 1993)
\bibitem{sagdeev}
R.Z. Sagdeev, Reviews of Plasma Physics (Consultants Bureau,
New York), {\bf{ 4}},  2391 (1966).
\bibitem{andersen 1968}
H.K. Andersen, N. D'Angelo, and P. Michelsen, Phys. Fluids {\bf{11}}, 606 (1968).
\bibitem{shukla2000}
P. K. Shukla, Phys. of Plasmas,  {\bf{7}(3)}, 1044-1046 (2000).
\bibitem{popel2000}
S. I. Popel, A. A. Gisko, A. P. Golub, T. V. Losseva, R. Bingham, and P. K. Shukla, Phys. of Plasmas, {\bf{7}}, 2410-2416 (2000).
\bibitem{shukla2001}
P. K. Shukla and A. A. Mamun, IEEE transaction on Plasma science, {\bf{29}}, 2 (2001).
\bibitem{hamid2009}
Hamid Reza Pakzad, and Kurosh Javidan, Pramana jour. of Phys. {\bf{73}}(5), 913-926 (2009).
\bibitem{samsonov2003}
D. Samsonov, G. Morfill, H. Thomas, T. Hagl, and H. Rothermel, V. Fortov, A. Lipaev, V. Molotkov, A. Nefedov, and O. Petrov, A. Ivanov and S. Krikalev,  Phys. Rev. E {\bf{67}}, 036404 (2003).
\bibitem{fortov2005}
V. E. Fortov, O. F. Petrov, V. I. Molotkov, M. Y. Poustylnik, V. M. Torchinsky, V. N. Naumkin, and A. G. Khrapak, Phys. Rev. E {\bf{71}}, 036413 (2005).
\bibitem{heinrich2009}
J. Heinrich, S.-H Kim, and R.L. Merlino, Phys. Rev. Lett. {\bf{103}}, 115002 (2009).
\bibitem{nakamura2012}
Y. Saitou, Y. Nakamura, T. Kamimura and O. Ishihara, Phys. Rev. Lett. {\bf{108}}, 065004 (2012).
\bibitem{usachev2014}
A Usachev, A Zobnin, O Petrov, V Fortov, M H Thoma, H H$\ddot{o}$fner, M Fink, A Ivlev and G Morfill, New Journal of Physics, {\bf{16}}, 053028 (2014).
\bibitem{surabhi2015}
S. Jaiswal, P. Bandyopadhyay, A. Sen, Rev. Sci. Instrum., {\bf{86}}(11), 113503 (2015).
\bibitem{thompsion}
C. Thompson, A. Barkan, N. DÕAngelo, and R.L. Merlino, Phys. Plasmas 4, 2331 (1997).
\bibitem{jyoti2002}
J. Pramanik, G. Prasad, A. Sen, and P. K. Kaw Phys. Rev. Lett. \textbf{88}, 175001 (2002).
\bibitem{merlino_1999}
Q. -Z. Luo, N. D' Angelo, and R. L. Merlino, Phys. Plasmas, \textbf{6}, 3455 (1999).
\bibitem{pakzad_2011}
H. R. Pakzad, Shock Waves, \textbf{21}, 357 (2011). 
\bibitem{annibaldi2007}
S V Annibaldi, A V Ivlev, U Konopka, S Ratynskaia, H M Thomas, G E Morfill, A M Lipaev, V I Molotkov, O F Petrov and V E Fortov, New J. Phys. \textbf{9}, 327 (2007).
\bibitem{kaw1998}
P. K. Kaw and A.Sen, Phys. Plasmas {\bf{5}}, 3552 (1998).
\bibitem{ichimaru1987}
S. Ichimaru, H. Iyetomi and S. Tanaka, Phys. Rep. \textbf{149}, 91 (1987).
\bibitem{nakamura2002}
Y. Nakamura, phys. Plasmas, \textbf{9}, 440 (2002).
\bibitem{epstein1924}
P. Epstein, Phys. Rev. \textbf{23}, 710 (1924).



\end{thebibliography}
\end{document}